\documentclass{article}
\usepackage{graphicx}  
\usepackage{amsmath}   
\usepackage[compress]{cite}
\usepackage{amssymb}   
\usepackage{bm} 
\usepackage{dcolumn}
\usepackage{color}
\usepackage{mathrsfs}
\usepackage{amsfonts}
\usepackage{varioref}
\usepackage[caption=false]{subfig}
\RequirePackage[colorlinks,citecolor=blue,urlcolor=magenta,linkcolor=blue]{hyperref}
\input epsf
\labelformat{section}{Section #1} 
\labelformat{subsection}{Section #1} 
\labelformat{subsubsection}{Section #1}
\labelformat{subsubsubsection}{Section #1}
\labelformat{equation}{Eq.~(#1)} 
\labelformat{figure}{Fig.~#1} 
\labelformat{subfigure}{Fig.~\thefigure#1} 
\labelformat{table}{Tab.~#1} 
\labelformat{appendix}{Appendix #1}
\title{In quest of axionic hairs in quasars}
\author{Indrani Banerjee
\footnote{tpib@iacs.res.in},
Bhaswati Mandal
\footnote{tpbm3@iacs.res.in},
and Soumitra SenGupta
\footnote{tpssg@iacs.res.in}\\
{\small{Department of Theoretical Physics, Indian Association for the Cultivation of Science, Kolkata-700032, India}}}

\begin{document}
  
\maketitle
\begin{abstract}
The presence of axionic field can provide plausible explanation to several long standing problems in physics such as dark matter and dark energy. The pseudo-scalar axion whose derivative corresponds to the Hodge dual of the Kalb-Ramond field strength in four dimensions plays crucial roles in explaining several astrophysical and cosmological observations. Therefore, the detection of axionic hairs/Kalb-Ramond field which appears as closed string excitations in the heterotic string spectrum may provide a profound insight to our understanding of the current universe. 
The current level of precision achieved in solar-system based tests employed to test general relativity, is not sufficient to detect the presence of axion. However, the near horizon regime of quasars where the curvature effects are maximum seems to be a natural laboratory to probe such additions to the matter sector. 
The continuum spectrum emitted from the accretion disk around quasars encapsulates the imprints of the background spacetime and hence acts as a storehouse of information regarding the nature of gravitational interaction in extreme situations. The surfeit of data available in the electromagnetic domain provides a further motivation to explore such systems. Using the optical data for eighty Palomar Green quasars we demonstrate
that the theoretical estimates of optical luminosity explain the observations best when the axionic field is assumed to be absent. However, axion which violates the energy condition seems to be favored by observations which has several interesting consequences. Error estimators, including reduced $\chi^{2}$, Nash-Sutcliffe efficiency, index of agreement and modified versions of the last two are used to solidify our conclusion and the implications of our result are discussed.

\end{abstract}
\section{Introduction}\label{Accretion_Intro}
General Relativity (GR), the most successful candidate in explaining the space-time structure around us, has exhibited remarkable agreement with a host of experimental tests \cite{Will:2005yc,Will:1993ns,Will:2005va,Yunes2013}. Yet, the inadequacy of GR to explain observations like galactic rotation curves and accelerated expansion of the universe \cite{CLIFTON20121,0004-637X-517-2-565,1538-3881-116-3-1009} indicates that either some additional matter fields are required or the gravity sector itself needs to be modified. This gave birth to a multitude of alternative theories, e.g., f(R) gravity \cite{PhysRevD.68.123512,NOJIRI201159,CAPOZZIELLO2006135}, Lanczos-Lovelock gravity \cite{doi:10.1063/1.1665613,Lanczos:1932zz,PADMANABHAN2013115}, extra-dimensional models \cite{PhysRevD.62.024012,PhysRevD.69.064020,PhysRevD.74.104031}, string cosmology \cite{McAllister2008} and Horndeski/scalar-tensor theories \cite{PhysRevLett.112.251102,0264-9381-33-15-154002} which can potentially fulfill the deficiencies in GR.  
However, one needs to test these theories against cosmological, astrophysical and solar system based observations so that they can be constrained/established/falsified.
Therefore, efforts to examine GR/alternate theories have been extended from weak-field solar system based observations to stronger gravitational fields involved in black hole accretion \cite{Armitage:2004ga}, binary compact objects\cite{vanStraten:2001zk} and cosmology \cite{Hwang:2001pu}. The electromagnetic emission from the accretion disk around black holes \cite{Chakrabarti:2004uz,0264-9381-34-11-115003,0004-637X-843-1-25}, precise radio observations and measurement of the rate of orbital decay of binary pulsars with subsequent emission of gravitational waves \cite{Hulse:1974eb,Stairs2003} and fluctuations in the cosmic microwave background (CMB) spectra\cite{EspositoFarese:1999pa,Damour:1996xx,CLIFTON20121}, all encode the imprints of the nature background space-time and hence can act as potential probes to test GR or other alternate theories.

Placed against this scenario, in this work, we aim to modify the matter sector with the pseudo-scalar field axion and study its impact on the continuum spectrum from the quasars. The motivation for 
considering such a field in the theory are many. In four space-time dimensions, 
the derivative of the axionic field can be identified as the Hodge-dual of the Kalb-Ramond field strength $H_{\mu \nu \alpha}$, a third rank completely anti-symmetric tensor field, which  
appears in the effective low energy action of a type IIB string theory \cite{Kalb:1974yc,Green:1987mn}. The field strength $H_{\mu \nu \alpha}$ corresponds to the massless, second rank anti-symmetric Kalb-Ramond field $B_{\mu \nu}$ that plays a crucial role in explaining several astrophysical and cosmological scenarios, e.g., Kalb-Ramond field giving rise to topological defects which might lead to intrinsic angular momentum of the structures in our galaxies \cite{0264-9381-12-2-016}, Kalb-Ramond field affecting the cosmic microwave background anisotropy \cite{1475-7516-2004-06-005}, Kalb-Ramond field being instrumental in understanding leptogenesis \cite{ELLIS2013407}, gravity theories formulated using twistors necessitates Kalb-Ramond field \cite{HOWE199680}, Kalb-Ramond field generating optical activity in spacetime \cite{0264-9381-19-4-304} to name a few. Therefore, the detection of axionic hairs/Kalb-Ramond field in astrophysical/ cosmological observations has far-reaching consequences.

A further impetus to 
study the nature and consequences of the Kalb-Ramond field is provided by superstring theory \cite{Kalb:1974yc,Green:1987mn}, where the Kalb-Ramond field strength $H_{\mu \nu \alpha}$,  
is found to bear a striking resemblance with space-time torsion \cite{Majumdar:1999jd,PhysRevD.81.024021,Sur:2016bwu,0264-9381-12-2-016,Hehl:1976kj,deSabbata:1994wi,Capozziello:2001mq,Lue:1998mq}. Torsion tensor $T^\alpha_{\mu\nu}$ in general relativity, is a third rank tensor associated with the anti-symmetric part of the affine connection, i.e., $T^\alpha_{\mu\nu}=\Gamma^\alpha_{\mu\nu}-\Gamma^\alpha_{\nu\mu}$. It is generally anti-symmetric in two indices, however, its identification with the  Kalb-Ramond field strength becomes relevant only when we consider a special sub-class of the  torsion tensor antisymmetrized in all the three indices. Thus modifying the gravity theory with a completely anti-symmetric space-time torsion or introducing the Kalb-Ramond field in the matter sector are equivalent. 

The attempts to detect axionic hairs in solar system based tests reveals that the change incurred in the bending of light/perihilion precession of Mercury due to the presence of axion is far below the known error bars and hence, such effects cannot be detected by the present level of precision in the solar system tests\cite{Kar:2002xa}.  
Incidentally, it has been found that the continuum spectrum emitted from the accretion disk around quasars favors certain classes of alternate gravity theories, e.g., extra dimensions, Einstein Gauss-Bonnet gravity in higher dimensions and scalar hairs in black hole spacetime inherited by Horndeski models \cite{Banerjee:2017hzw}. Hence, in this paper we aim to examine if the continuum spectrum emitted by the quasars provide a suitable astrophysical laboratory to uncover the presence of axionic hairs.

\par
The paper is broadly classified into five sections. In \ref{KR}, we discuss the modifications introduced in the Einstein's equations due to the presence of Kalb-Ramond field and its equivalence with space-time torsion is established. The static, spherically symmetric and asymptotically flat solution of the field equations in such a scenario is also reviewed.
In \ref{Accretion_general}, we examine the spectrum emitted from the accretion disk around a black hole in the presence of Kalb-Ramond field. In \ref{result} we compare and analyze the theoretically obtained spectra and luminosities with the observed data. Finally, we conclude with a discussion of our results with some scope for future work in \ref{Accretion_Conc} . 

Throughout the paper, the Greek indices have been used to label the four dimensional spacetime indices. The metric convention adopted is (-,+,+,+).

\section{Static spherically symmetric black hole solutions in a Kalb-Ramond background}\label{KR}
Kalb-Ramond field, which transforms as an antisymmetric second rank tensor appears naturally in field theory and also in the heterotic string spectrum \cite{Kalb:1974yc, Green:1987mn}. Mathematically, it can be thought of as a generalization of the electromagnetic potential with two indices instead of one \cite{Kalb:1974yc,Majumdar:1999jd}. In this case, the gauge vector field $A_\mu$ in electrodynamics is replaced by a rank-2 antisymmetric tensor field $B_{\mu \nu}$, which is associated with a rank-3 antisymmetric field strength, $H_{\mu \nu \alpha}$. The field strength $H_{\mu \nu \alpha}$ corresponding to the Kalb-Ramond field $B_{\mu \nu}$, is given by $H_{\mu \nu \alpha}=\partial_{[\mu}B_{\nu\alpha]}=\frac{1}{3}[\nabla_\mu B_{\nu\alpha}+\nabla_\nu B_{\alpha\mu}+\nabla_\alpha B_{\mu\nu}]=\frac{1}{3}[\partial_\mu B_{\nu\alpha}+\partial_\nu B_{\alpha\mu}+\partial_\alpha B_{\mu\nu}]$. Analogous to electrodynamics, the Lagrangian density of the Kalb-Ramond field is represented by the square of the field strength.

The full action for Kalb-Ramond field in 4 dimensional Einstein gravity is given by, 
\begin{equation}
\mathcal{S} = \int d^{4}x\sqrt{-g}\bigg[\frac{\bar{R}}{16\pi G} - \frac{1}{12}H_{\alpha\beta\gamma}H^{\alpha\beta\gamma}\bigg] \label{Eq1}
\end{equation}
where $G$ is the four dimensional gravitational constant and $\bar{R}$, the Ricci scalar.
Since Kalb-Ramond field is a second rank anti-symmetric tensor it should have six independent components in four dimensions. However, only the spatial components of the Kalb-Ramond field are dynamical which reduces the propagating degrees of freedom from six to three. On account of the additional gauge symmetry present in the system, $B_{\mu\nu} \rightarrow B_{\mu\nu} + \nabla_{\mu}\chi_{\nu} - \nabla_{\nu}\chi_{\mu}$, the number of degrees of freedom reduces to zero, since the gauge field $\chi_{\nu}$ has 3 spatial components. Now, by replacing the gauge field as, $\chi_{\nu} \rightarrow \chi_{\nu} + \partial_{\nu}\psi$, where $\psi$ is a scalar degree of freedom, it results in the Kalb-Ramond field to have a single degree of freedom in 4 spacetime dimensions. The factor of $-1/12$ in the Lagrangian corresponding to the Kalb-Ramond field ensures that the kinetic term in local inertial frame appears as $\frac{1}{2}(\partial_{t}B_{\mu \nu})^2$ in our metric convention. 

\par
In order to obtain the gravitational field equations the action given by \ref{Eq1} is varied with respect to the metric. The resulting field equations are, 
\begin{align}
G_{\mu \nu} = 8 \pi G T_{\mu\nu}^{(KR)}
\end{align}
In general, the energy momentum tensor $T_{\mu\nu}$, corresponding to a Lagrangian density $\mathcal{L}$ which may be due to matter or any arbitrary field  is defined by,
\begin{align}
T_{\mu\nu} = -\frac{2}{\sqrt{-g}}\frac{\delta(\sqrt{-g}\mathcal{L})}{\delta g^{\mu\nu}}
\end{align}
Thus, the energy momentum tensor corresponding to the Kalb-Ramond field is given by,
\begin{align}
T_{\mu\nu}^{KR} = \frac{1}{6}\bigg[3H_{\mu\rho\sigma}H_{\nu}^{\rho\sigma} - \frac{1}{2}\big\{H_{\rho\sigma\delta}H^{\rho\sigma\delta}\big\}g_{\mu\nu}\bigg]
\end{align}

Varying the action with respect to $B_{\alpha\beta}$ leads to the corresponding field equations for Kalb-Ramond field, $\nabla_{\mu}H^{\mu\nu\rho} = 0$. It can also be shown that Kalb-Ramond field satisfies the Bianchi identity, $\nabla_{[\mu}H_{\alpha\beta\gamma]} = 0$. 
The fact that the massless, Kalb-Ramond field has a single degree of freedom in 4 dimensions indicates that we can express it in terms of a pseudo-scalar field $\Phi$, where, 
\begin{align}
H^{\mu\nu\rho} = \epsilon^{\mu\nu\rho\sigma}\partial_{\sigma}\Phi=\partial^{[\mu}B^{\nu\rho]}\label{Eq5}
\end{align}
In fact, the field strength $H^{\mu\nu\rho}$ is the hodge dual of the derivative of the pseudo-scalar field $\Phi$, also known as axion. \ref{Eq5} establishes the identification of the Kalb-Ramond field with the axionic field and hence throughout the paper we will use the terms Kalb-Ramond field and axion synonymously.

Next, we seek for a static, spherically symmetric solution of the Einstein's equations in presence of Kalb-Ramond field.
The corresponding line element is given by,
\begin{align}
ds^{2} = -f(r)dt^{2} + \frac{dr^{2}}{g(r)} + r^{2} d\Omega^{2} \label{line}
\end{align}
where, $f(r)$ and $g(r)$ are arbitrary functions of the radial coordinate.
It is important to note that the three form, $H_{\mu \nu \alpha}$ has four non-zero components $H_{012}$, $H_{013}$, $H_{023}$ and $H_{123}$. 
However, the requirement of static, spherical symmetric solution, which satisfies asymptotic flatness ensures that only, $H_{023} \neq 0$ and all other components vanish
\cite{Kar:2002xa}. Thus, the axionic field turns out to be a function of only radial coordinates.
Hence, $H^{023} = \epsilon^{0231}\Phi^{\prime}(r)$, where `prime' denotes the derivative with respect to the radial coordinate. The non-trivial energy-momentum tensor components are thus,
\begin{align}
T_{0}^{0 (KR)} = \frac{1}{2}H^{023}H_{023} = -h(r)^{2} \\
T_{1} ^{1 (KR)} = -\frac{1}{2}H^{023}H_{023}= h(r)^{2} = -T_{2} ^{2 (KR)}=-T_{3} ^{3 (KR)}
\end{align}
The static, spherically symmetric, asymptotically flat solution of the Einstein field equations in presence of Kalb-Ramond field is already studied in detail in Kar et al. and Chakraborty \& SenGupta \cite{Kar:2002xa,Chakraborty:2017uku}. Hence, we simply quote the results here,
\begin{align}
f(r) = 1 - \frac{2}{r} + \frac{b}{3r^{3}} + \mathcal{O}(\frac{1}{r^4}) \label{fr} 
\end{align}
\begin{align}
g(r) = 1- \frac{2}{r} + \frac{b}{r^{2}} +  \mathcal{O}(\frac{1}{r^3}) \label{gr}.
\end{align}
Here, the distance $r$ is expressed in units of $M$ and $b$ represents the axionic hair associated with the black hole solution and has units of $M^2$. It is important to note that in order to preserve the energy condition, $b$ should be positive. Also $|b|$ is less than 1, since the metric coefficients involve a perturbative expansion in $b$ \cite{Kar:2002xa}. Thus the line element given by \ref{line} with metric components given by \ref{fr} and \ref{gr} can be thought of as a perturbation over the Schwarzschild solution due to the presence of Kalb-Ramond field.

\section{A General Discussion on Accretion Disk in Presence of Kalb-Ramond field}\label{Accretion_general}
In order to probe the effect of Kalb-Ramond field or its dual axion in the strong gravity regime around quasars, we concentrate on the electromagnetic emission from the  surrounding accretion disk. The nature of the electromagnetic spectrum depends both on the background spacetime and also on the structure and properties of the accretion disk. As already discussed, in this work we consider the background spacetime to be static, spherically symmetric and asymptotically flat generated by a blackhole with axionic hair such that the metric ansatz is given by, \ref{line},\ref{fr},\ref{gr}.

We study the process of accretion in such a spacetime considering the thin disk approximation \cite{Shakura:1972te,Novikov:1973kta}.
Under this approximation, we assume that the accretion disk is geometrically thin, such that $z(r)/r\ll 1$ ($z(r)$ being the vertical thickness of the disk at a radial distance $r$) and optically thick such that the disk emits locally as a black body \cite{Shakura:1972te,Novikov:1973kta}. 
The central plane of the accretion disk is assumed to coincide with the equatorial plane of the black hole. The gas of the disk is assumed to maintain almost circular orbits with an additional negligible radial velocity arising due to viscous stresses, which facilitates infall of matter into the black hole. Further, the vertical motion is considered to be negligibly small and the disk is assumed to accrete at a steady rate
i.e., local/turbulent behavior is restricted to a length scale $\sim z(r)$ and time scale $\sim \mathcal{O}$ (time interval required for the accreting gas to travel a radial distance equivalent to z(r)) which does not affect the global behavior of the physical quantities.

The energy momentum tensor associated with the accreting fluid has contributions from: 
a) the geodesic flow $\rho u^\alpha u^\beta$; 
b) the specific internal energy of the system $\tilde{\Pi}$; 
c) the stress-tensor as measured in local rest frame of the accreting fluid $\tilde{t}^{\alpha\beta}$ and 
d) the energy flux $\tilde{q}^\alpha$. 
The corresponding expression for stress-energy tensor takes the form,
\begin{align}
\tilde{T}^{\mu\nu}=\rho(1+\tilde{\Pi})u^\mu u^\nu+\tilde{t}^{\mu\nu}+u^\mu \tilde{q}^\nu+u^\nu \tilde{q}^\mu
\end{align}
The average of the $(0,z)$ component of $\tilde{T}^{\mu\nu}$, which essentially corresponds to $\langle \tilde{q}^z\rangle$, gives rise to the flux of radiation coming out of this system. 
The outgoing photons undergo repeated collisions with the accreting fluid such that matter and radiation attain an equlibrium and the disk emits locally as a blackbody. 
Hence the disk is geometrically thin but optically thick. Consequently, Stefan Boltzmann Law holds at every radial distance $r$, i.e., $T_{eff}(r) = (F(r)/\sigma)^{1/4}$, where $F(r)$ is the flux from the disk at $r$. In other words, each small annulus of the disk emits a Planck spectrum at a different peak temperature given by $T_{eff}(r)$. The collective emission from the disk which is an envelope of all these black body spectra is known as the multi-color black body/multi temperature black body spectrum which generally peaks in the optical band for supermassive black holes. This nature of spectrum is a consequence of the thin disk approximation of the accretion model. The nature of the background spacetime affects $T_{eff}(r)$ through its dependence on $F(r)$.

The accreting fluid is assumed to obey conservation of mass, angular momentum and energy. These are coupled differential equations which simplify considerably in the thin disk approximation where one assumes that the disk truncates at the marginally stable circular orbit $r_{ms}$, across which no viscous stress can act. Subsequently, the flux $F(r)$ emanated from the accretion disk can be written in a closed analytic form given by,

\begin{align}
F(r) = \frac{3G\dot{m}M}{8\pi R^{3}}\frac{\mathcal{Q}}{\mathcal{B\sqrt{C}}} \label{flux}
\end{align} 
where, 
\begin{align}
\mathcal{B} = \sqrt{\frac{2}{r^{2}f^{\prime}}} \label{B}
\end{align}
\begin{align}
\mathcal{C} = \frac{(2f - rf^{\prime})}{r^{2}f^{\prime}} \label{C}
\end{align}
for a spherically symmetric metric. The expression of $\mathcal{Q}$ is given by,
\begin{align}
\mathcal{Q}=\mathcal{L}-\frac{3}{2}\sqrt{\frac{1}{r}}\mathcal{I}\int _{r_{\rm ms}}^{r}dr~\frac{\mathcal{L}}{\mathcal{B}\mathcal{C}\mathcal{I}}\sqrt{\frac{1}{r^{3}}} \\
\mathcal{L} = \frac{1}{\sqrt{\mathcal{C}}}-\frac{L_{\rm ms}}{\sqrt{r}} \\
\mathcal{I} = \exp\left[\frac{3}{2}\int _{r}^{\infty}dr~\frac{1}{\mathcal{B}\mathcal{C}r^{2}} \right]
\end{align}
$L_{ms}$ is the angular momentum corresponding to the marginally stable circular orbit, $\dot{m}$ is the steady state accretion rate and $R=rr_g$ is the dimensionful radial distance, $r_g=GM/c^2$ being the gravitational radius. 
The metric dependence of the the emitted flux appears through the quantities $\mathcal{B}$, $\mathcal{C}$, $\mathcal{Q}$ and through the radius of the marginally stable circular orbit $r_{ms}$.
The innermost stable circular orbit $r_{ms}$ corresponds to the inflection points of the effective potential $V_{eff}(r)$, which in the case of a spherically symmetric metric is obtained by solving,
\begin{equation}\label{Acc_Eq_09}
2rf(r)f''(r)-4rf(r)'^{2}+6f(r)f'(r)=0
\end{equation}
 
For a detailed discussion on the radial structure of the accretion disk in the thin disk approximation, one is referred to \cite{Novikov:1973kta,Page:1974he,Banerjee:2017hzw}.
It is important to note at this stage that the metric dependence on the emission from the accretion disk comes only from the $g_{tt}$ component of the spherically symmetric metric (f(r) in our case) and the form of the $g_{rr}$ component is irrelevant here \cite{Banerjee:2017hzw}. The role of $g_{rr}$ gains prominence when we try to investigate the nature of the event horizon which is obtained by solving $g^{rr}=0$.

For our problem, the form of $f(r)$ and $g(r)$ are respectively given by \ref{fr} and \ref{gr}. Hence,
\begin{align}
\mathcal{B} = \bigg[1 - \frac{b}{2r^{2}}\bigg]^{-1/2} \\
\mathcal{C} = \bigg[1 - \frac{3}{r} + \frac{5}{6}\frac{b}{r^{3}}\bigg]\bigg(1 - \frac{b}{2r^{2}}\bigg)^{-1}
\end{align}
We can obtain $r_{ms}$ by solving \ref{Acc_Eq_09} which turns out to be a function of $b$.

Thus, the expression for the flux takes the form,
\begin{align}\label{flux}
F(r,b) = \frac{3G\dot{m}M}{8\pi{r_g}^3 r^{3}}&\frac{1 - \frac{b}{2r^{2}}}{\sqrt{1 - \frac{3}{r} + \frac{5}{6}\frac{b}{r^{3}}}}\Bigg\{\sqrt{\frac{1 - \frac{b}{2r^{2}}}{1 - \frac{3}{r} + \frac{5}{6}\frac{b}{r^{3}}}} - \frac{L_{ms}}{\sqrt{r}} 
\nonumber
\\
& - \frac{3}{2\sqrt{r}}\exp\Bigg[-\frac{3}{2}\int_{r}^{\infty}\frac{d\bar{r}}{\bar{r}^{2}}\frac{(1 - \frac{b}{2\bar{r}^{2}})^{3/2}}{1 - \frac{3}{\bar{r}} + \frac{5}{6}\frac{b}{\bar{r}^{3}}}\Bigg]\int^{r}_{r_{ms}}\frac{dr^{\prime}}{r^{\prime 3/2}}\Bigg(\sqrt{\frac{1 - \frac{b}{2r^{\prime 2}}}{1 - \frac{3}{r^{\prime}} + \frac{5}{6}\frac{b}{r^{\prime 3}}}} - \frac{L_{ms}}{\sqrt{r^{\prime}}}\Bigg)
\nonumber
\\
&\times\exp\Bigg[\frac{3}{2}\int_{{r^{\prime}}}^{\infty}\frac{dr^{\prime \prime}}{r^{\prime \prime 2}}\frac{(1 - \frac{b}{2r^{\prime 2}})^{3/2}}{1 - \frac{3}{r^{\prime \prime}} + \frac{5}{6}\frac{b}{r^{\prime \prime 3}}}\Bigg]\frac{(1 - \frac{b}{2r^{\prime 2}})^{3/2}}{1 - \frac{3}{r^{\prime}} + \frac{5}{6}\frac{b}{r^{\prime 3}}}\Bigg\}
\end{align}

Consequently, $T_{eff}(r,b)=\big[F(r,b)/\sigma\big]^{1/4}$, is given by,
\begin{align}\label{T}
T_{eff}(r,b) = \Bigg[\frac{3G\dot{m}M}{8\sigma\pi{r_g}^3 r^{3}}&\frac{1 - \frac{b}{2r^{2}}}{\sqrt{1 - \frac{3}{r} + \frac{5}{6}\frac{b}{r^{3}}}}\Bigg\{\sqrt{\frac{1 - \frac{b}{2r^{2}}}{1 - \frac{3}{r} + \frac{5}{6}\frac{b}{r^{3}}}} - \frac{L_{ms}}{\sqrt{r}} 
\nonumber
\\
& - \frac{3}{2\sqrt{r}}\exp\Bigg[-\frac{3}{2}\int_{r}^{\infty}\frac{d\bar{r}}{\bar{r}^{2}}\frac{(1 - \frac{b}{2\bar{r}^{2}})^{3/2}}{1 - \frac{3}{\bar{r}} + \frac{5}{6}\frac{b}{\bar{r}^{3}}}\Bigg]\int^{r}_{r_{ms}}\frac{dr^{\prime}}{r^{\prime 3/2}}\Bigg(\sqrt{\frac{1 - \frac{b}{2r^{\prime 2}}}{1 - \frac{3}{r^{\prime}} + \frac{5}{6}\frac{b}{r^{\prime 3}}}} - \frac{L_{ms}}{\sqrt{r^{\prime}}}\Bigg)
\nonumber
\\
&\times\exp\Bigg[\frac{3}{2}\int_{{r^{\prime}}}^{\infty}\frac{dr^{\prime \prime}}{r^{\prime \prime 2}}\frac{(1 - \frac{b}{2r^{\prime 2}})^{3/2}}{1 - \frac{3}{r^{\prime \prime}} + \frac{5}{6}\frac{b}{r^{\prime \prime 3}}}\Bigg]\frac{(1 - \frac{b}{2r^{\prime 2}})^{3/2}}{1 - \frac{3}{r^{\prime}} + \frac{5}{6}\frac{b}{r^{\prime 3}}}\Bigg\}\Bigg]^{1/4}
\end{align}

Since each annulus of the disk emits a Planck spectrum given by
\begin{align}
B_\nu(T_{eff})=\frac{2h\nu^3}{c^2}\frac{1}{e^{\frac{h\nu}{kT_{eff}}}-1} 
\end{align}
where, $T_{eff}$ is given by \ref{T}, the luminosity $L_\nu$ from the disk is obtained by integrating the Planck function $B_\nu(T_{eff})$ over the disk surface. Thus $L_\nu$ assumes the form,
\begin{align}
L_\nu&=r_g^2\cos i\int_0^{4\pi}d\Omega   \int_{r_{\rm in}}^{r_{\rm out}} \int_0^{2\pi}d\phi \sqrt{-g}r drB_{\nu}(T_{\rm eff})\nonumber \\ 
&=8\pi^2 r_g^2\cos i \int_{r_{\rm in}}^{r_{\rm out}} \sqrt{-g}B_{\nu}(T_{\rm eff})r dr
\label{L_nu}
\end{align}
where, $L_\nu$ is the luminosity emitted over $4\pi$ solid angle at frequency $\nu$ , $i$ is the angle of inclination of the disk to the line of sight and $g$ is the determinant of the metric. The variation of $\nu L_\nu$ with $\nu$ gives the spectrum from the thin disk for a particular value of $b$. This is plotted in \ref{Fig_01}.
Since the emission from the thin disk around supermassive black holes generally peak in the optical part of the spectrum we will be interested in comparing our theoretical estimates of optical luminosity with observations. 

Theoretically derived optical luminosity is obtained by the relation,
\begin{align}\label{L_cal}
L_{cal} = \nu L_{\nu}
\end{align}
where, $\nu$ represents the frequency corresponding to the wavelength $4861\AA$ \cite{Davis:2010uq}. 

In order to compare the calculated optical luminosity $L_{cal}$ with observations, a group of eighty Palomar Green quasars are considered whose optical and bolometric luminosities are reported \cite{Davis:2010uq}. The masses of these quasars have also been constrained by the technique of reverberation mapping and in some cases $M-\sigma$ relation is also employed to estimate the mass. Since we are working with quasars, the inclination angle $i$ is believed to lie in the range $0.5 < cos i < 1$. As nearly edge-on systems are likely to be obscured, a typical value of $cos i = 0.8$ is assumed. 
This is further substantiated by the fact that in the Schwarzschild scenario, the error (e.g., reduced $\chi^{2}$, Nash-Sutcliffe efficiency, index of agreement and the modified versions of the last two) between the observed and theoretical luminosities gets minimized when $cos i$ lies in the range $0.77-0.82$ \cite{Banerjee:2017hzw}. 

In the next section we will study how the presence of $b$ affects $L_{cal}$ and compare the analytically obtained results with observations.

\section{Results Based on Numerical Analysis}\label{result}

\begin{figure}[htp]
\centering
\includegraphics[scale=1]{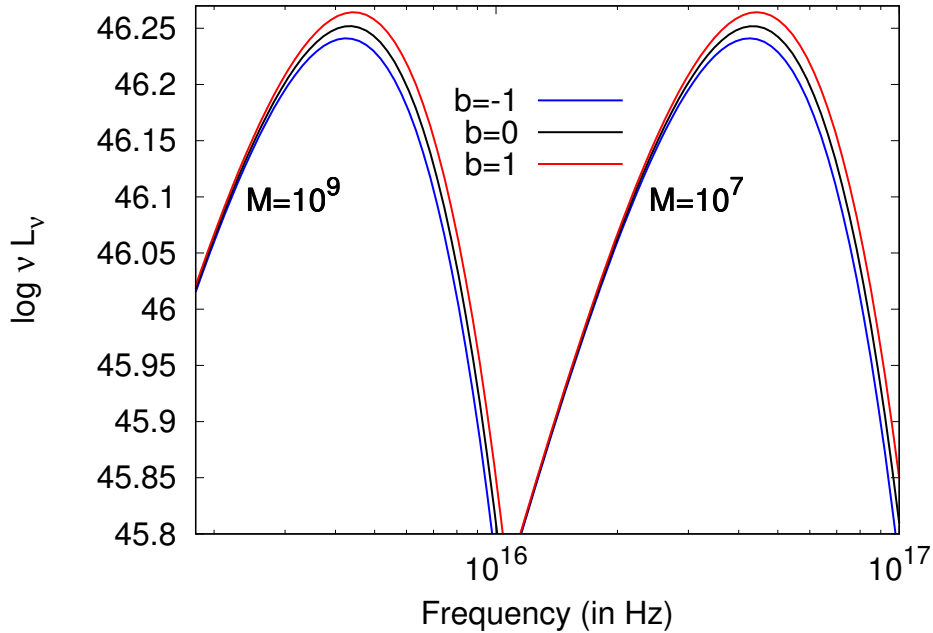}
\caption{The variation of the theoretically derived spectra from the accretion disk 
is illustrated with the axion parameter $b$ and compared with the Schwarzschild scenario, $b=0$. The effect of $b$ becomes somewhat pronounced only in the high frequency regime for both the masses of the central black hole $M=10^7 M_{\odot}$ and $M=10^9 M_{\odot}$. It is clear that a positive $b$ enhances the luminosity from the Schwarzschild value while a negative $b$ decreases the luminosity. The accretion rate assumed is $1 M_{\odot}\textrm{yr}^{-1}$ and $\cos i$ is taken to be 0.8. For further discussions see text.}
\label{Fig_01}
\end{figure}

\begin{figure}[htp]
\subfloat[Reduced $\chi^{2}$ \label{Fig_2a}]{\includegraphics[scale=0.5]{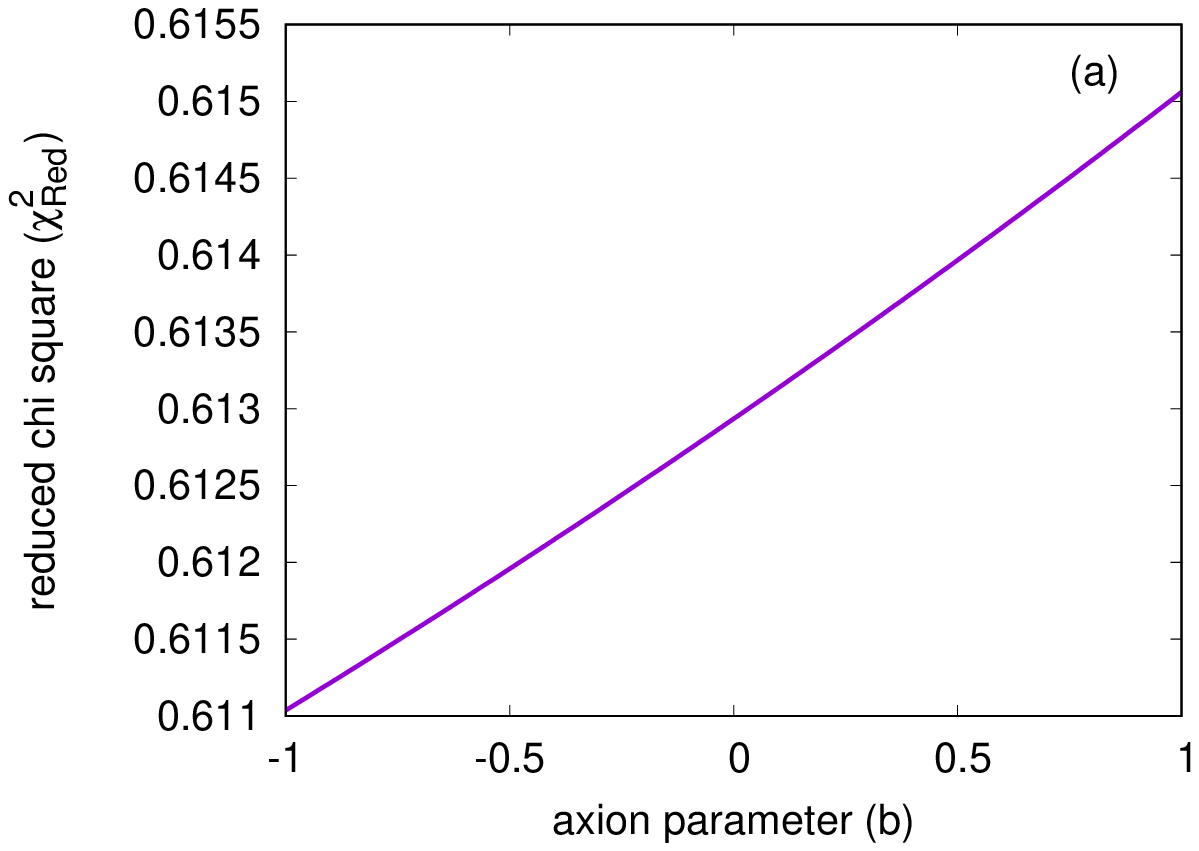}}
\subfloat[Nash-Sutcliffe efficiency \label{Fig_2b}]{\includegraphics[scale=0.5]{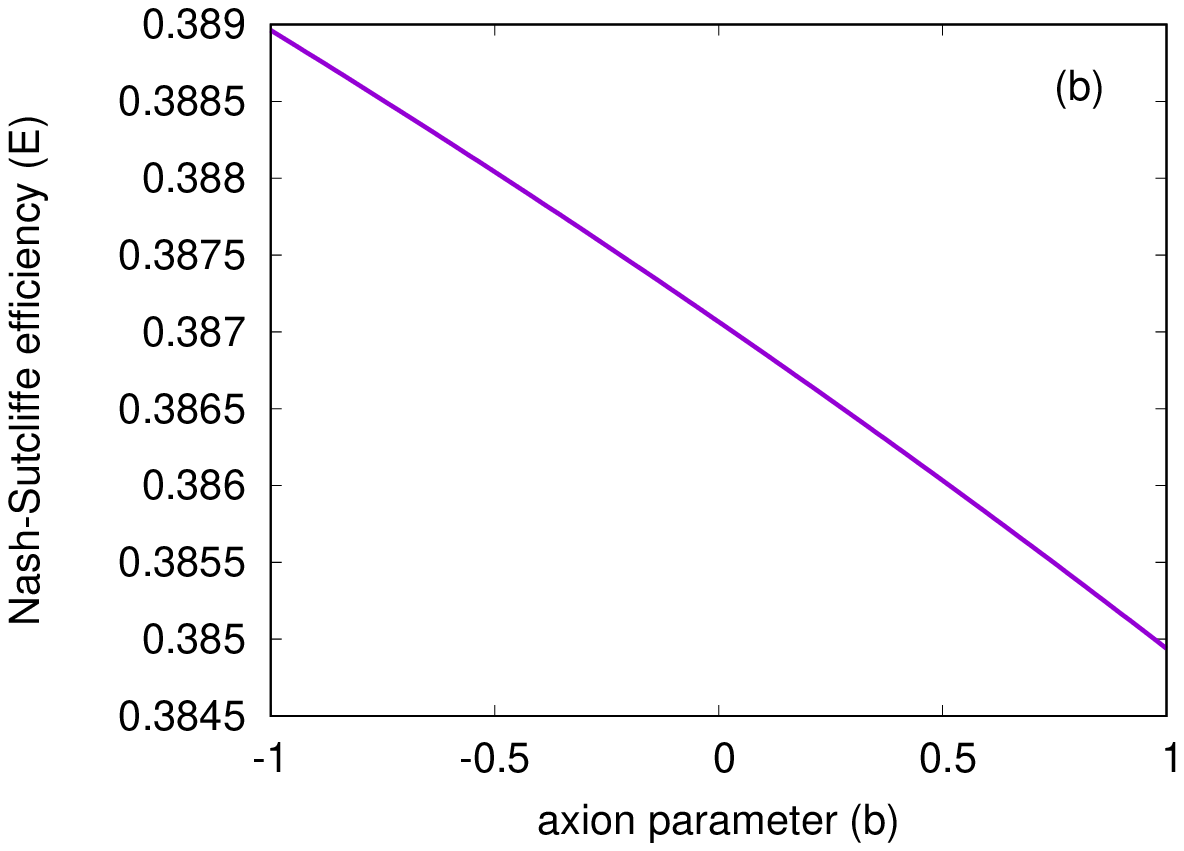}}
\subfloat[Modified Nash-Sutcliffe efficiency \label{Fig_2c}]{\includegraphics[scale=0.5]{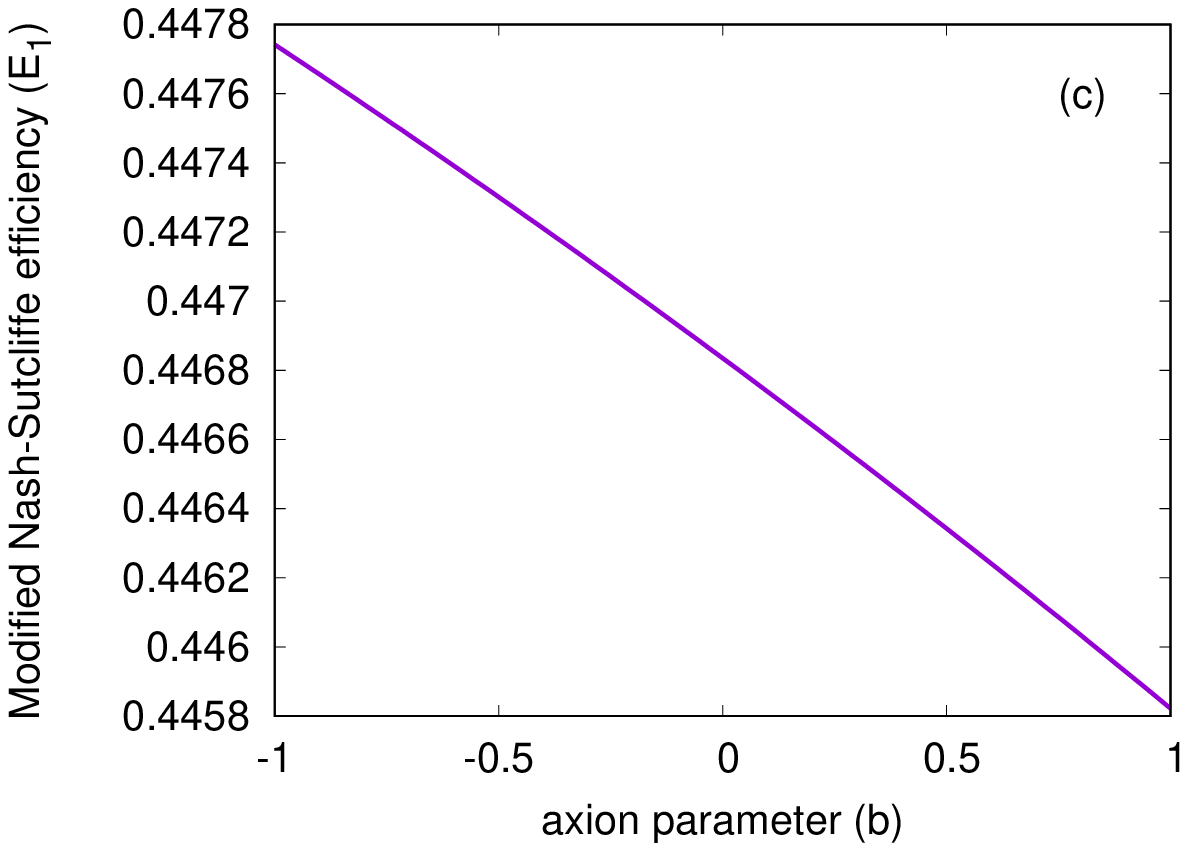}}
\caption{The above figure depicts variation of (a) the reduced $\chi^2$, $\chi^2_{Red}$, (b) the Nash-Sutcliffe efficiency $E$ and (c) the modified Nash-Sutcliffe efficiency $E_{1}$ with the axion parameter $b$.}
\label{Fig_02}
\end{figure}

\begin{figure}[htp]
\subfloat[Index of agreement \label{Fig_3a}]{\includegraphics[scale=0.65]{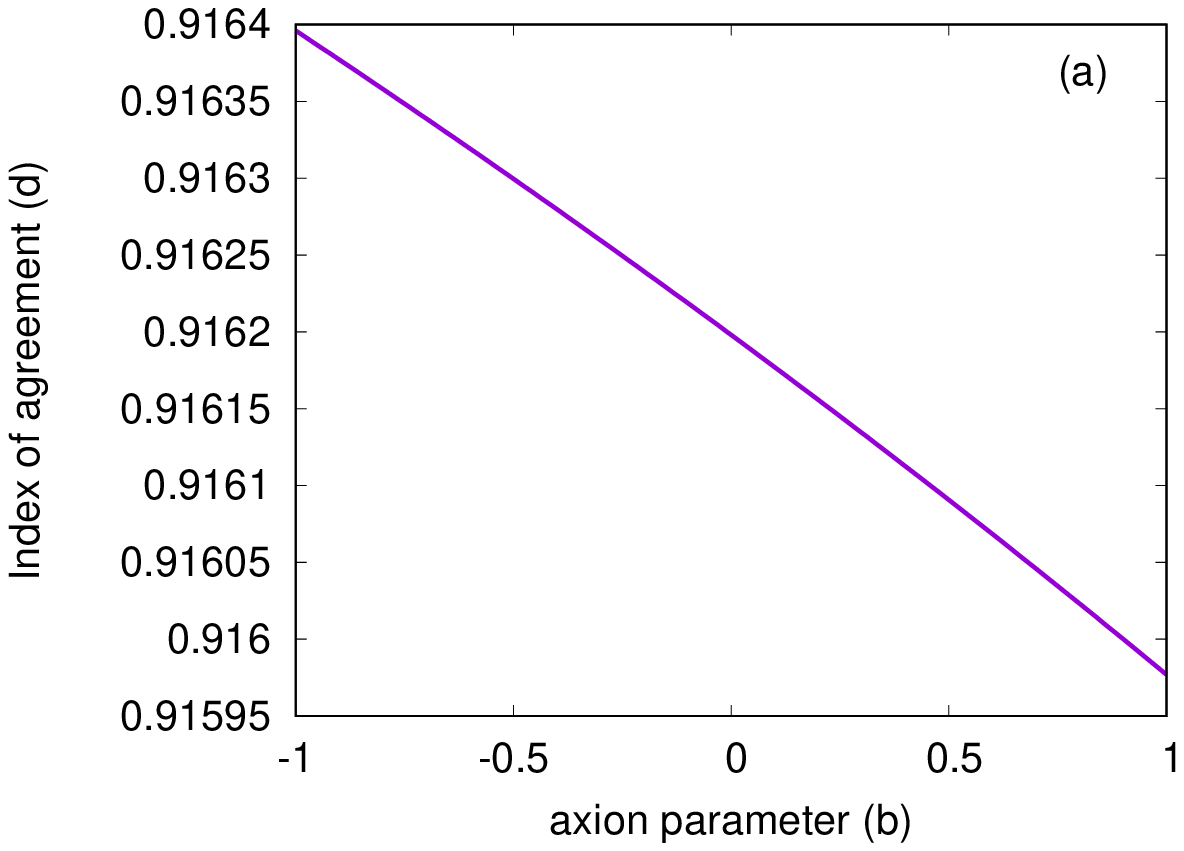}}
\subfloat[Modified Index of agreement\label{Fig_3b}]{\includegraphics[scale=0.65]{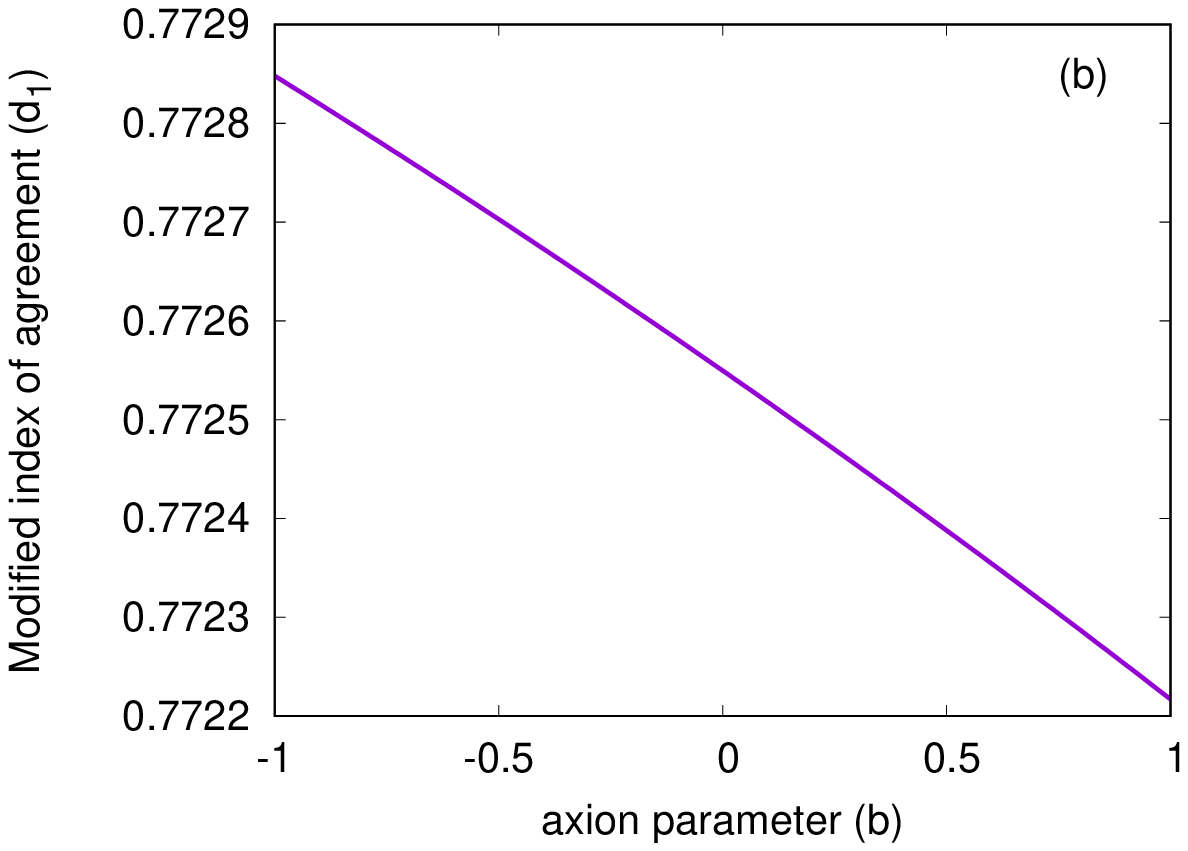}}
\caption{The above figure depicts variation of (a) the index of agreement $d$ and (b) the modified index of agreement $d_1$ with the axion parameter $b$.}
\label{Fig_03}
\end{figure}

In this section, the theoretical values of optical luminosity is derived for a sample of eighty PG quasars studied in \cite{Schmidt:1983hr,Davis:2010uq} using the thin accretion disk model described in \cite{Novikov:1973kta,Page:1974he}. The accretion is studied in the background of the space-time metric given by \ref{line} with the metric components given in \ref{fr} and \ref{gr}. The technique of reverberation mapping is used to constrain the masses of these PG quasars \cite{Kaspi:1999pz,Kaspi:2005wx,Boroson:1992cf,
1972ApJ...171..467B,1974ATsir.831....1L,1982ApJ...255..419B,Peterson:2004nu}. 
For thirteen quasars the $M-\sigma$ relation is also used to constrain their masses \cite{Ferrarese:2000se,Gebhardt:2000fk,Dasyra:2006jy,Wolf:2008sm,Tremaine:2002js}. 
Using emissions in the optical \cite{1987ApJS...63..615N}, UV \cite{Baskin:2004wn}, far-UV \cite{Scott:2004sv}, and soft X-ray \cite{Brandt:1999cm} the bolometric luminosity of these quasars are determined. The observed values of the optical luminosities and the accretion rates of these eighty PG quasars are reported in \cite{Davis:2010uq}.

In order to obtain a qualitative understanding of the effect of the axion/Kalb-Ramond field on the emission from the accretion disk, we derive the spectra emitted from the thin disk using \ref{L_nu}, where the flux is given by \ref{flux}.
\ref{Fig_01} illustrates the effect of a non-zero Kalb-Ramond field on the spectral emission from the accretion disk. The effect of Kalb-Ramond field becomes somewhat appreciable only in the high frequency regime, since that part of the spectra is emitted from the inner region of the accretion disk where the effect of the background metric becomes more pronounced. Since the term associated with Kalb-Ramond field only gives a $1/r^3$ correction to the Schwarzschild metric (See \ref{fr}) it does not modify the spectrum substantially even in the high energy domain.
This result stands true for both the masses $M=10^7 M_{\odot}$ and $M=10^9 M_{\odot}$ although the peak frequency is higher for a lower mass quasar since the peak temperature $T$ for a multi-color black body spectrum scales as $T \propto M^{-1/4}$ \cite{2002apa..book.....F}.

The above discussion illuminates that the effect of Kalb-Ramond field has a small but non-trivial effect on the emitted spectra of the quasars. We will quantify this statement further at the end of this section. 
In order to understand whether the presence of Kalb-Ramond field explains the observed spectra better, we calculate the optical luminosity $L_{cal}$ for each of these quasars at $\nu=4861 \AA$ \cite{Davis:2010uq} using the masses and accretion rates reported in \cite{Davis:2010uq}. We evaluate the optical luminosity for the entire set of eighty PG quasars using \ref{L_cal} assuming the average value of $b$ for the quasars varies from -1 to 1. Since we are considering a perturbative expansion in $b$ we consider $|b|<1$.
Next we evaluate several error estimates to achieve a more quantitative understanding of the most favored model of Kalb-Ramond field. 

\begin{itemize}

\item \textbf{Reduced}~$\mathbf{\chi^{2}}~$:~
The $\chi ^{2}$ of a distribution is defined as:
\begin{align}\label{Acc_Eq_21}
\chi ^{2}(b)=\sum _{i}\frac{\left\{\mathcal{O}_{i}-\mathcal{M}_{i}(b) \right\}^{2}}{\sigma _{i}^{2}}.
\end{align}
Here, $\{\mathcal{O}_{i}\}$ represents the set of observed data with possible errors $\{\sigma_{i}\}$  which provide proper weightage to each observations while $\{\mathcal{M}_{i}(b)\}$ represents the model estimates of the observed quantity which depends on the model parameter $b$. In our case, the errors $\sigma_{i}$ associated with the observed optical luminosity $L_{opt}$ are not reported since the systematic uncertainties in the estimation methods override the statistical uncertainty in the input data. Hence we assume that each observation has equal weightage. A more frequently used description of error is given by the reduced $\chi^2$ estimate, $\chi^{2}_{\rm Red}$, which is obtained by dividing the $\chi^2$ with the number of degrees of freedom. The value of $b$ for which $\chi^{2}_{\rm Red}$ is minimized corresponds to the most favorable value of $b$. Note that, $b$ here represents the average $b$ of all the eighty quasars. \ref{Fig_2a} elucidates the variation of the  $\chi^{2}_{\rm Red}$ with the axion parameter $b$. Since, $\chi^{2}_{\rm Red}$ exhibits a monotonically increasing behavior from $b=-1$ to $b=1$, it implies that a negative value of $b$ is favored. This indicates, Kalb-Ramond field violating the energy condition is favored by observations. Before discussing the implications and significance of this result we shall consider some more error estimates to confirm the authenticity of this result.

\item \textbf{Nash-Sutcliffe Efficiency:} 
Nash-Sutcliffe Efficiency \cite{NASH1970282,WRCR:WRCR8013,2005AdG.....5...89K} is related to the sum of the absolute squared differences between the predicted and the observed values normalized by the variance of the observed values. It is given by,
\begin{align}\label{Acc_Eq_22}
E(b)=1-\frac{\sum_{i}\left\{\mathcal{O}_{i}-\mathcal{M}_{i}(b)\right\}^{2}}{\sum _{i}\left\{\mathcal{O}_{i}-\mathcal{O}_{\rm av}\right\}^{2}}
\end{align}
where $\mathcal{O}_{\rm av}$ denotes average of the observed value of the optical luminosity from the accreting disk. It varies from $-\infty ~\rm to ~ 1$, where a model with $E<0$ indicates that the mean value of the observed data would have been a better predictor than the model. Unlike $\chi^{2}_{\rm Red}$, the model which maximizes $E$ is the most favored model. \ref{Fig_2b} illustrates the variation of $E$ with $b$. It shows that $E(b)$ monotonically decreases from $-1$ to $1$ emphasizing that a negative value of $b$ is favored from observations. Note however that the variation of $E(b)$ with $b$ is very small and the value of $E(b)\sim 0.38$ (which is true for our case) represents a model which is satisfactory although one can in principle look for better models. \cite{Goyal}. We will discuss more on this issue in the next section.

\item \textbf{Modified Nash-Sutcliffe Efficiency:}
Modified form of Nash-Sutcliffe efficiency $E_1$ is used to overcome
the oversensitivity of Nash-Sutcliffe efficiency to higher values of the optical luminosity \cite{WRCR:WRCR8013}. Such oversensitivity is introduced by the mean square error in the Nash-Sutcliffe efficiency. Thus, modified Nash-Sutcliffe efficiency which is defined as,
\begin{align}
E_{1}(b)&=1-\frac{\sum_{i}|\mathcal{O}_{i}-\mathcal{M}_{i}(b)|}{\sum _{i}|\mathcal{O}_{i}-\mathcal{O}_{\rm av}|}
\end{align}
is used to enhance the sensitivity of this estimator for lower values as well. A model which maximizes $E_1$ represents the most favored model.
The variation of $E_1$ with the average axion parameter $b$ is illustrated in \ref{Fig_2c}. Just like $E$, $E_1$ also decreases monotonically with increase in $b$ and the conclusions obtained from the previous two error estimates remain unaltered.

\item \textbf{Index of agreement and its modified form:}
The index of agreement, denoted by $d$ was proposed  \cite{willmott1984evaluation, doi:10.1080/02723646.1981.10642213,2005AdG.....5...89K} to overcome the insensitivity of Nash-Sutcliffe efficiency towards the differences between the observed and predicted means and variances \cite{WRCR:WRCR8013}. It is defined as:
\begin{align}\label{Acc_Eq_23}
d(b)=1-\frac{\sum_{i}\left\{\mathcal{O}_{i}-\mathcal{M}_{i}(b)\right\}^{2}}{\sum _{i}\left\{|\mathcal{O}_{i}-\mathcal{O}_{\rm av}|+|\mathcal{M}_{i}(b)-\mathcal{O}_{\rm av}|\right\}^{2}}
\end{align}
The quantity in the denominator is called the potential error which represents the
largest value the squared difference of each pair of observed and predicted values can attain. Since the denominator in $d$ is larger compared to $E$ for every pair, the index of agreement is always greater than the corresponding Nash-Sutcliffe efficiency.

Index of agreement $d$ also suffers from the same drawback as the Nash-Sutcliffe efficiency i.e., its oversensitivity to greater values of the optical luminosity
because of the presence of a square term in the numerator \cite{WRCR:WRCR8013} and hence a similar modified estimator is used. The modified index of agreement $d_1$ is given by,
\begin{align}\label{Acc_Eq_24}
d_{1}(b)&=1-\frac{\sum_{i}|\mathcal{O}_{i}-\mathcal{M}_{i}(b)|}{\sum _{i}\left\{|\mathcal{O}_{i}-\mathcal{O}_{\rm av}|+|\mathcal{M}_{i}(b)-\mathcal{O}_{\rm av}| \right\}}
\end{align}
The model which maximizes $d$ and $d_1$ is considered to be a better model. 
\ref{Fig_3a} and \ref{Fig_3b} elucidates the variation of $d$ and $d_1$ with $b$ which essentially replicates the trend exhibited by $E$ and $E_1$ in \ref{Fig_2b} and \ref{Fig_2c} respectively.

It is important to note that the presence of axionic hairs in black holes appears only as a perturbation to the Schwarzschild metric. This prevents the choice of $|b|>1$. Further, the spectrum from the accretion disk is affected mainly by the $g_{tt}$ component of the metric (in our case f(r) given by \ref{fr}), where the leading order term associated with the Kalb-Ramond field has a $1/r^3$ suppression. Thus the effect of Kalb-Ramond field on the emitted spectra is hardly perceptible and hence the variation in the values of the error estimators is very small as b is increased from -1 to 1. This can be easily verified from \ref{Fig_02} and \ref{Fig_03}. 
Nevertheless, the result obtained is important and non-trivial
because the entire analysis discussed above corroborates the fact that astrophysical observations associated with quasar optical data signals the absence of Kalb-Ramond field/axionic hairs in quasars. However, it seems that Kalb-Ramond field  which violates the energy condition, i.e., negative $b$, explains the observations better.
The efficacy of such a scenario has been discussed in different context such as the bouncing model of the universe to avoid big bang singularity \cite{PhysRevD.77.044030},  removal of singularity in geodesic congruences\cite{Kar:2006ms},
Buchdahl's limit for star formation \cite{Chakraborty:2017uku} and possible source of a spacetime with non-zero cosmological constant inherited from the bulk Kalb-Ramond field in a higher dimensional scenario \cite{CHAKRABORTY2016258}, where such energy violating term appears in an effective field theory due to the presence of an anti-symmetric tensor field.

\end{itemize}

\section{Concluding Remarks}\label{Accretion_Conc}
In this work we aim to discern the effect of the Kalb-Ramond field from the continuum spectrum emitted by the accretion disk around quasars. 
The presence of axionic field produce negligible effects on the solar system based tests such as perihelion precession of mercury and bending of light and hence cannot be detected within the experimental precision. 
This is because, if the existence of axion changes the bending of light and perihelion precession of mercury within the observational error bars, it would lead to an extraordinarily high Kalb-Ramond energy density (since the leading order correction to the metric due to the Kalb-Ramond field falls as $1/r^3$ which is hardly detectable at large length scales) that would lead to a huge ambient background temperature. Since this is not observed, we conclude that the change induced in the bending of light and perihelion precession  due to the presence of Kalb-Ramond field is much lower than 
the known error bars and hence the present level of precision in the solar system tests
cannot detect their presence.

Thus, it is instructive to investigate what imprints it has in the strong gravity regime around quasars. The continuum spectrum emitted by the accretion disk around black holes is intricately related to the properties of the background space-time and hence it can be exploited to constrain/establish/invalidate several modified gravity theories. 
The plethora of electromagnetic data available from the quasar observations provides a further motivation to explore such systems. 
The scope for probing such strong gravity regimes will be further enhanced with the advent of the Event Horizon Telescope \cite{PhysRevLett.116.031101} which aims at imaging the event horizon of a black hole using the techniques of Very Large Baseline Interferometry (VLBI) which can further provide strong constraints on several alternative gravity theories.
 
Kalb-Ramond field manifests itself as a perturbation in the Schwarzschild metric when static, spherically symmetric and asymptotically flat solutions of the gravitational field equations are examined. Since the leading order term associated with the Kalb-Ramond field causes a $1/r^3$ correction to the Schwarzschild solution, its effects on the spectrum is hardly conspicuous. 
Nevertheless, we find that it does give us an indication whether quasar observations favors/discards the presence of such a field.
By studying the impact of such a perturbed metric on the continuum spectrum we derive the optical luminosity emitted by eighty Palomar Green quasars in the thin-disk approximation and compare it with the optical data. We perform several error estimates e.g., reduced $\chi^{2}$, Nash-Sutcliffe efficiency, index of agreement and modified versions of the last two to constrain the axion/Kalb-Ramond field parameter 
in terms of the observational data. Interestingly, such an analysis reveals that even in the strong gravity regime around supermassive black holes, the presence of Kalb-Ramond field seems to be disfavored. Suppression of Kalb-Ramond field strength has been discussed by 
Mukhopadhyaya, Sen \& SenGupta \cite{Mukhopadhyaya:2002jn} in a warped brane-world scenario \cite{Randall:1999ee} with bulk Kalb-Ramond field. 
Further work by Das, Mukhopadhyaya \& SenGupta \cite{Das:2014asa} advocates that in such a higher dimensional scenario, the back reaction ensuing from the radius stabilization mechanism \cite{Goldberger:1999uk} can cause additional suppression of the Kalb-Ramond field on the effective four-dimensional visible universe. 

In addition to the above finding, our analysis also unfolds that axion/Kalb-Ramond field which disregards the energy condition is favored by electromagnetic observations from quasars! Such a scenario gains prominence in the context of bouncing cosmology, removal of singularity in geodesic congruences and plays a vital role in modifying the
Buchdahl's limit for star formation.

The current work presents a first step to unearth the effects of axion/Kalb-Ramond field from quasar data in the electromagnetic sector. Several extensions of this work are possible. Since, supermassive black holes in quasars can be rotating the continuum emission from the quasars needs to be studied in an axi-symmetric background in presence of Kalb-Ramond field. Further, the Novikov-Thorne model works only in the thin-disk approximation and fails to explain the full spectral energy distribution of quasars and hence a more improved accretion model needs to be considered. The assumption that the accretion disk extends upto the marginally stable circular orbit need not be correct. These effects should be eventually studied to determine the deviations they may incur on the conclusions of our current analysis.
Efforts are being made to incorporate a more comprehensive accretion model in the background of a spinning black hole in presence of Kalb-Ramond field and will be reported elsewhere.

\section*{Acknowledgements}
The research of SSG is partially supported by the Science and Engineering
Research Board-Extra Mural Research Grant No. (EMR/2017/001372), Government of India.
The authors thank Sumanta Chakraborty for helpful discussions and insightful comments throughout the course of this work. 

\bibliography{axion}

\bibliographystyle{./utphys1}

\end{document}